\documentclass[conference, 9pt]{IEEEtran}
\IEEEoverridecommandlockouts
\usepackage{cite}
\usepackage{amsmath,amssymb,amsfonts}
\usepackage{algorithmic}
\usepackage{graphicx}
\usepackage{textcomp}
\usepackage{wrapfig}
\usepackage{lipsum}
\usepackage{graphicx}
\usepackage{subcaption}
\usepackage[font=small,skip=2pt]{caption} % Reduce space around captions

\usepackage{threeparttable}
\usepackage{multirow}
\usepackage{multicol}
\usepackage{subcaption} % Add this in your preamble
\usepackage{amsmath}
\usepackage{circledtext}
\usepackage{tikz}
\usepackage{url}
\usepackage{graphicx}
\usepackage{float}
\usepackage{placeins}

\usepackage{xcolor}
\def\BibTeX{{\rm B\kern-.05em{\sc i\kern-.025em b}\kern-.08em
    T\kern-.1667em\lower.7ex\hbox{E}\kern-.125emX}}

\IEEEpubid{\makebox[\columnwidth]{ 979-8-3315-2710-5/25/\$31.00 ©2025 IEEE
\hfill{ \hspace{\columnsep} \makebox[\columnwidth]{ }}}}    
\begin{document}

\title{Unicorn-CIM: \underline{U}ncovering the Vul\underline{n}erability and \underline{I}mproving the Resilien\underline{c}e \underline{o}f High-P\underline{r}ecisio\underline{n} \underline{C}ompute-\underline{i}n-\underline{M}emory\\

\author{
    Qiufeng Li, Yiwen Liang, Weidong Cao \\
    Department of Electrical and Computer Engineering, The George Washington University, DC, USA \\
    \{qiufeng.li, yiwen.liang, weidong.cao\}@gwu.edu
}

% \title{Uncovering the Resilience of High-Precision Compute-in-Memory: Assessment and Enhancement\\
%A Resilient Framework for High-Precision Compute-in-Memory: Assessment and Enhancement\
}

% N in one
% N-in-1 (N2one, one4N)
% 1/NECC

% \and
% \and

\maketitle

\begin{abstract}
Compute-in-memory (CIM) architecture has been widely explored to address the von Neumann bottleneck in accelerating deep neural networks (DNNs).
However, its reliability remains largely understudied, particularly in the emerging domain of floating-point (FP) CIM, which is crucial for speeding up high-precision inference and on-device training.
This paper introduces Unicorn-CIM, a framework to \underline{\textbf{u}}ncover the vul\underline{\textbf{n}}erability and \underline{\textbf{i}}mprove the resilien\underline{\textbf{c}}e \underline{\textbf{o}}f high-p\underline{\textbf{r}}ecisio\underline{\textbf{n}} \underline{\textbf{CIM}}, built on static random-access memory (SRAM)-based FP CIM architecture.
Through the development of fault injection and extensive characterizations across multiple DNNs, Unicorn-CIM reveals how soft errors manifest in FP operations and impact overall model performance. 
Specifically, we find that high-precision DNNs are extremely sensitive to errors in the exponent part of FP numbers.
Building on this insight, Unicorn-CIM develops an efficient algorithm-hardware co-design method that optimizes model exponent distribution through fine-tuning and incorporates a lightweight Error Correcting Code (ECC) scheme to safeguard high-precision DNNs on FP CIM.
Comprehensive experiments show that our approach introduces just an $8.98\%$ minimal logic overhead on the exponent processing path while providing robust error protection and maintaining model accuracy.
This work paves the way for developing more reliable and efficient CIM hardware.

\end{abstract}

\iffalse
\begin{IEEEkeywords}
CIM, resilience, floating-point
\end{IEEEkeywords}
\fi

\section{Introduction}

Compute-in-memory (CIM) has been extensively studied as a promising solution to alleviate the longstanding von Neumann bottleneck in accelerating deep neural networks (DNNs).
Although significant strides have been made to improve the speed and energy efficiency of CIM macros \cite{survey, intcimv1, intcimv2, intcimv3}, these improvements often assume error-free memory operations -- a premise that does not hold in practice due to inherent limitations in current CIM macro designs.
Specifically, unlike commercial memory devices used for storage, which incorporate robust error detection and protection techniques, state-of-the-art CIM macros \cite{ cim2025_2, fpcim2024_2} lack strong safeguards against typical memory errors, such as soft (transient) and hard (permanent) errors.
Additionally, the widespread use of voltage scaling \cite{voltage_scale} to boost energy efficiency in CIM macros further exacerbates soft errors.
These hardware non-idealities markedly degrade the accuracy of DNNs when deployed on CIM.
Thus, recent studies\cite{etcim, mac-ecc, successive} have focused on improving the reliability of integer (INT) CIM macros by developing efficient error correction codes (ECCs). 

%over the past decade

Despite the tangible advances in the reliability enhancement of INT CIM macros, the reliability of floating-point (FP) CIM macros remains largely unexamined.
It is important to note that FP CIM has recently gained considerable attention to accelerate high-precision DNNs, fueled by the growing demand for high-accuracy inference and on-device training at the edge\cite{fpcim2023, fpcim2024_1, fpcim2024_2, fpcim2022}.
Numerous FP CIM chips \cite{cim2025_2, fpcim2023, fpcim2024_1, fpcim2024_2} have been implemented, demonstrating extraordinary energy efficiency in speeding up FP8/FP16 DNN models for both inference and training applications.
However, FP CIM is inherently more complex than INT CIM due to its involvement of exponent summation and mantissa multiplication (elaborated in Section \ref{FP arithmetic}).
As a result, neither the unreliability characterization frameworks nor the ECC methods developed for INT CIM macros\cite{etcim, mac-ecc, successive} can be readily applied to FP CIM macros.
\textbf{This gap underscores the pressing need for dedicated research to understand the impact of memory non-idealities on the accuracy of high-precision DNNs and to develop tailored solutions to enhance the resilience of FP CIM}.
Addressing this gap is essential to deploying trustworthy CIM systems on the edge for ubiquitous real-world DNN applications \cite{dnn_appl}, such as embodied AI and autonomous driving.

This work proposes \textbf{Unicorn-CIM}, a comprehensive framework designed to \underline{\textbf{u}}ncover the vul\underline{\textbf{n}}erability and \underline{\textbf{i}}mprove the resilien\underline{\textbf{c}}e \underline{\textbf{o}}f high-p\underline{\textbf{r}}ecisio\underline{\textbf{n}} \underline{\textbf{CIM}}.
Specifically, we focus on static random-access memory (SRAM)-based FP CIM architecture for in-depth study, as it is the most widely used practical implementation\cite{intcimv1, intcimv2, intcimv3}  at the edge to accelerate high-precision inference and training.
We begin with fault injection and extensive characterizations to examine how memory bit errors manifest in FP operations and affect the overall performance of various DNNs. 
Our findings reveal that high-precision DNNs are particularly sensitive to errors in the exponent part of FP numbers. 
Building on this insight, we propose a novel algorithm-hardware co-design method to safeguard high-precision DNNs on FP CIM.
This co-design optimizes model exponent distribution through fine-tuning (algorithm optimization) and incorporates a tailored lightweight ECC scheme in CIM macro development (hardware optimization).
Through co-optimization, our approach significantly reduces hardware overhead while maintaining robust error protection and preserving DNN accuracy.
Our work advances the development of more dependable and efficient CIM hardware that sustains the increasingly sophisticated demands of edge learning applications.
Key contributions of this work include:

\begin{itemize}

\item \textbf{Comprehensive unreliability characterization}: We build a fault-injection framework for FP DNNs and conduct a thorough examination of how memory errors affect the accuracy of various models, providing valuable insights into error propagation and the development of safeguard techniques for FP CIM.

\item \textbf{Lightweight algorithm-hardware co-optimization}: We tailor a lightweight ECC scheme for FP DNNs through efficient algorithm-hardware co-design, effectively addressing the unique challenges inherent in the current design of FP CIM macros.

\item \textbf{Significant resilience improvement with minimal resource overhead}: Extensive evaluations show that our co-design incurs minimal logic overhead while providing robust error protection and maintaining model accuracy.

\end{itemize}

\section{Background and Related Work}

\subsection{INT CIM Macros: Error Sources and Protection Techniques}
%\textcolor{red}{Discuss the different errors in CIM with NVM and VM. Hard errors and soft errors.}

\subsubsection{Primary memory error resources} CIM macros built on volatile and non-volatile memory (NVM) suffer from various device non-idealities, leading to memory errors and threatening reliability.
The primary error sources are hard stuck-at faults, which arise from permanent defects in chip fabrication, and soft device variations that result in transient bit flips during read/write, both ultimately causing inaccuracies in multiplication-and-accumulation (MAC) operations.  
While digital CIM macros avoid errors along the computing path, they must still contend with hard and soft errors in memory devices.
For example, a study on 14 nm SRAM\cite{ber_sram} shows that soft bit errors (i.e., bit error rate (BER)) increase exponentially with lower supply voltages (Fig.~\ref{CIM}(a)). 
For analog CIM, peripheral components such as analog-to-digital converters (ADCs) also contribute to dynamic inaccuracies \cite{mac-ecc}, further exacerbating system error rates. 
As a result, DNN inference accuracy can significantly degrade if adequate protection for the CIM macros is not applied.
Since the occurrence probability of soft errors is significantly higher, especially under low voltage conditions, and they tend to recur, whereas hard errors are relatively rare, easier to resolve, and compatible with error correction circuits designed for soft errors\cite{etcim, soc_soft_test}, our study focus is primarily on soft error mitigation.

\subsubsection{{State-off-the-art protection methods}}
Prior studies have explored MAC protections in INT CIM macros. 
For instance, \cite{etcim} implemented block-wise ECC in digital SRAM CIM. 

Recent analog CIM macros have also applied column repair or bit-wise ECC.
In \cite{secded}, the traditional Single Error Correction Double Error Detection (SECDED) ECC, commonly applied in commercial memory systems, is adapted for protection.
Yet, this approach is effective only under a relatively low BER. 
To enhance error correction capability, \cite{successive} combines successive correction with SECDED-CIM\cite{secded}, enabling double and triple error correction following error detection.
While these protection methods are efficient and low-latency, none can readily apply to FP CIM for high-precision DNNs due to the unique computation formats of FP numbers, as introduced in the next section.
%Section \ref{limitaiom}.

\subsection{\textbf{FP CIM Macros: Primers and  Challenges}}
\label{fp_operation}

\begin{figure}[!t]
\centering
\includegraphics[width=1.0\linewidth]{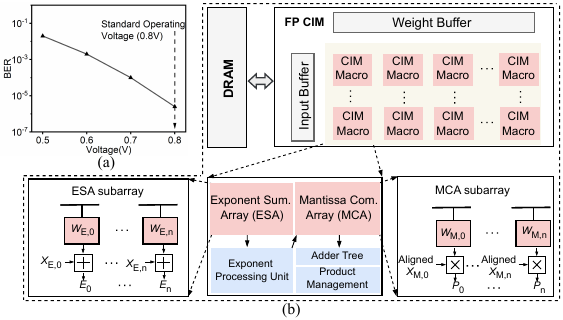}
\caption{(a) BER vs power supply for SRAM under 14 nm technology \cite{ber_sram}.  (b) A general FP CIM architecture used in current research.}
\label{CIM}
\vskip -12pt
\end{figure}

\subsubsection{{Primers of FP arithmetic and FP CIM}}
\label{FP arithmetic}
Compared to a signed integer with only two parts, i.e., sign bit and magnitude, an FP number consists of three parts: sign (S), exponent (E), and mantissa (M).
Taking the FP16 format, widely used by current FP CIM macro designs\cite{fpcim2023, fpcim2024_1, fpcim2024_2, cao2024addition, cao_pim, yi, fpcim2022}, as an example, these parts are 1-b, 5-b, and 10-b. 
The additional exponent component makes FP MAC computations more complex than INT MAC operations, as detailed below.

\noindent\textbf{FP Multiplication:} FP multiplication involves exponent summation (step \tikz[baseline=(char.base)]{
    \node[shape=circle, draw, inner sep=0.5pt] (char) {1};
}) and mantissa multiplication (step \tikz[baseline=(char.base)]{
    \node[shape=circle, draw, inner sep=0.5pt] (char) {2};
}).
The process is illustrated in Eq.~\eqref{eq:fp_mult} using two positive numbers for simplicity. Note that the ‘1’ preceding 'M' is a hidden bit, which is not explicitly represented in the binary format of an FP number.
\begin{equation}
    (2^{E_0} \cdot 1.M_0) \cdot (2^{E_1} \cdot 1.M_1) = \underbrace{2^{E_0+E_1}}_{\tikz[baseline=(char.base)]{
    \node[shape=circle, draw, inner sep=0.5pt] (char) {1};
}} \cdot \underbrace{(1.M_0 \cdot 1.M_1)}_{\tikz[baseline=(char.base)]{
    \node[shape=circle, draw, inner sep=0.5pt] (char) {2};
}}.
\label{eq:fp_mult}
\end{equation}
\textbf{FP Addition:} The addition of FP numbers is non-trivial and is expressed in Eq.~\eqref{eq:fp_addition}. 
The process begins with alignment, which involves finding the maximum exponent (step \tikz[baseline=(char.base)]{
    \node[shape=circle, draw, inner sep=0.5pt] (char) {1};
}), \( E_{\text{max}} = \max\{E_0, E_1\} \), among all exponents  and computing the exponent difference, \( E_{0(1)} - E_{\text{max}} \), between each exponent and the maximum exponent (step \tikz[baseline=(char.base)]{
    \node[shape=circle, draw, inner sep=0.5pt] (char) {2};
}). Next, the mantissa part is shifted to the right according to the exponent difference and then summed (step \tikz[baseline=(char.base)]{
    \node[shape=circle, draw, inner sep=0.5pt] (char) {3};
}). 
Finally, the sum undergoes additional processing, such as truncation, to conform to the standard FP format.
% \begin{equation} \label{eq:fp_addition}
%     2^{E_0} \cdot 1.M_0 + 2^{E_1} \cdot 1.M_1 = \underbrace{2^{E_{\max}}}_{\tikz[baseline=(char.base)]{
%     \node[shape=circle, draw, inner sep=0.5pt] (char) {1};
% }} \cdot 
%     \left\{
%     \begin{array}{c}
%         \underbrace{2^{E_0 - E_{\max}} \cdot 1.M_0}_{\tikz[baseline=(char.base)]{
%     \node[shape=circle, draw, inner sep=0.5pt] (char) {2};
% }} \\[10pt]
%         + \\[10pt]
%         \underbrace{2^{E_1 - E_{\max}} \cdot 1.M_1}_{\tikz[baseline=(char.base)]{
%     \node[shape=circle, draw, inner sep=0.5pt] (char) {2};
% }}
%     \end{array}
%     \right\}_{\tikz[baseline=(char.base)]{
%     \node[shape=circle, draw, inner sep=0.5pt] (char) {3};
% }}.
% \end{equation}
\begin{equation}
2^{E_0}\cdot1.M_0+2^{E_1}\cdot1.M_1 = \underset{\tikz[baseline=(char.base)]{
    \node[shape=circle, draw, inner sep=0.5pt] (char) {1};
}}{\underbrace{ 2^{{E_{\max}}}}} \cdot \underset{\tikz[baseline=(char.base)]{
    \node[shape=circle, draw, inner sep=0.5pt] (char) {3};
}}{\underbrace{
\left\{
            \begin{array}{lr}
              \underset{\tikz[baseline=(char.base)]{
    \node[shape=circle, draw, inner sep=0.5pt] (char) {2};
}}{\underbrace{2^{{E_{0}}-{E_{\max}}}}} \cdot 1.M_{0} &  \\
                +&  \\
              \underset{\tikz[baseline=(char.base)]{
    \node[shape=circle, draw, inner sep=0.5pt] (char) {2};
}}{\underbrace{2^{{E_{1}}-{E_{\max}}}}} \cdot 1.M_{1}
             \end{array}
\right.}}.
    \label{eq:fp_addition}
\end{equation}

\noindent{\textbf{FP MAC:}}  By generalizing Eq.~\eqref{eq:fp_addition} to the accumulation of $n$ FP numbers, where each number, $2^{E_i}\cdot 1.M_{i}$ is assumed to be a product of a weight-activation pair, i.e., $(-1)^{W_{S,i}}\cdot2^{W_{E,i}}\cdot 1.W_{M,i}$ and $(-1)^{X_{S,i}}\cdot 2^{X_{E,i}}\cdot 1.X_{M,i}$ based on Eq.~\eqref{eq:fp_mult} (i.e., $E_i=W_{E,i}+X_{E,i}$ and $1.M_{i}=1.W_{M,i}\cdot1.X_{M,i}$), the MAC of FP numbers is achieved. 
%This operation is fundamental to FP DNN models.
FP MAC enables high-precision computations in DNN models, supporting the highest accuracy and best training quality.
%Thus, efficient hardware acceleration for FP VMM is highly desirable.

\noindent\textbf{FP CIM:} 
Recently, FP CIM has been widely studied to accelerate high-precision DNNs that rely on FP MAC operations, driven by the demand for high-accuracy inference and on-device training\cite{fpcim2023, fpcim2024_1, fpcim2024_2, fpcim2022}.
Current FP CIM implementations largely rely on digital computations with SRAM\cite{fpcim2023, fpcim2024_1, fpcim2024_2, fpcim2022, sun}, primarily due to their exceptional reliability and endurance compared to NVM.
While there are non-trivial differences at the micro-architecture level, these implementations commonly share the same system architecture and main components. 
Major components at the macro level include an Exponent Summation Array and an Exponent Processing Unit, which are responsible for operations such as exponent summation, identifying the maximum exponent, and aligning the mantissa based on exponent differences, as well as a Mantissa Computing Array, Adder Trees, and Product Management, which are responsible for mantissa multiplication, as shown in Fig.~\ref{CIM}(b).
These components are tailored to perform essential processing steps in FP MAC as expressed by Eqs.~\eqref{eq:fp_mult}  and~\eqref{eq:fp_addition}.
Additionally, other peripheral modules, such as input/output buffers and controllers, are included to complete the data flow. 
Our co-design builds on this general architecture with SRAM as the hardware substrate.
Specifically, we aim to use minimal hardware resources to protect FP MAC as detailed in Section~\ref{sec: framework}.

\subsubsection{{Challenges to tackle the unreliability of FP CIM}}
The complex FP MAC expressed in Eqs.~\eqref{eq:fp_mult} and~\eqref{eq:fp_addition} shows that the different parts in an FP weight could have different impacts on the model accuracy. 
A number of fault injection frameworks have been developed to characterize the resilience of DNNs on INT CIM. 
As an example, Ares \cite{ares} introduced a framework capable of supporting
%both static and dynamic 
fault injection for quantized DNNs.
%, validating its model with real hardware measurements. 
Yet, such frameworks cannot be applied to high-precision DNNs.
This presents a challenge to studying the unreliability of FP CIM.
We develop a fault injection framework (Section~\ref{sec: characterization}) to bridge this gap to facilitate static and dynamic bit-flip injections in FP DNNs.
While prior research has examined the impact of bit-flip errors in FP numbers on the control logic of computing systems~\cite{fpu_2012}, no studies have been focused on how these bit-flip errors affect the computation process and, ultimately, the accuracy of DNNs deployed in CIM.

\section{Unicorn-CIM Framework}
\label{sec: framework}

We first build a fault injection framework to assess the resilience of high-precision DNNs on non-ideal SRAM FP CIM that suffers from inherent bit-flip errors.
Based on characterizations, we propose our novel co-design approach to enhance the robustness of FP CIM in accelerating high-precision DNN applications.

\begin{figure}[!t]
\centering
\includegraphics[width=1.0\linewidth]{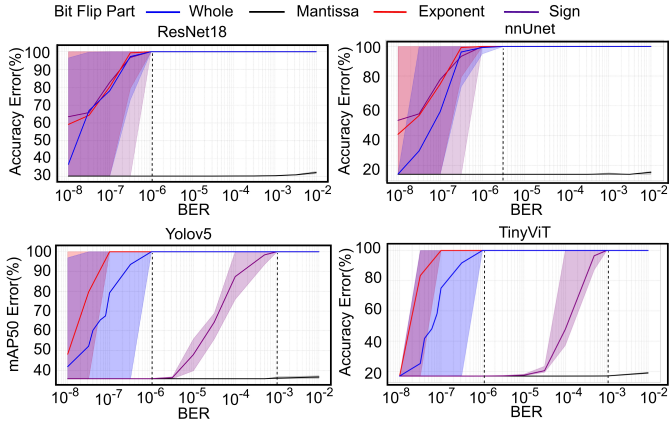}
\caption{Inference accuracy by high-precision DNNs vs. BER.}
\vskip -6pt
\label{fig:square_images}
\end{figure}

\subsection{Resilience of FP DNNs: Characterizations and Insights}
\label{sec: characterization}

We use a benchmark of commonly employed DNNs (Table~\ref{models}) in edge applications for resilience characterizations.
This benchmark includes tasks such as image classification, object detection, and object segmentation, as well as various network architectures, including transformer-based and convolutional architectures.
All models are pre-trained with FP16. 

We build a fault injection framework to characterize the resilience of these FP DNN models to memory bit flips.
We employ two ways for fault injection: static injection and dynamic injection.  
Static injection is used for inference, where faults are injected into the stationary weights for once.
This simulates the scenario where the model is deployed on CIM for static inference tasks.  
Dynamic injection, on the other hand, is used for training, where faults are injected during runtime as weights are frequently accessed. 
This simulates the scenario where the model is deployed on CIM for training. 
Using our customized fault injection framework, we introduce random bit flips across various parts of the FP weights, including the sign, exponent, mantissa, or full value. 
This approach enables us to accurately assess the sensitivity of model accuracy to each component of the FP weights. 
For each BER, we conduct $100$ independent runs on each model to ensure reliable inference accuracy measurements, resulting in a total of $24,000$ experiments.

\subsubsection{{Characterizations}}
We found that the training process is completely disrupted even with small BERs (e.g., $1 \times 10^{-6}$, see Fig.~\ref{train}), making it difficult to assess the sensitivity of each component in FP DNNs to bit-flip errors in macro memory. 
As a result, we present the characterization results on the benchmark with inference applications (Fig.~\ref{fig:square_images}).
Notably, we find that FP DNNs are highly sensitive to bit errors on the exponent field: 
model accuracy begins to fluctuate significantly at a BER of $1 \times 10^{-8}$ and degrades to zero at a BER of $1 \times 10^{-6}$.
In contrast, FP DNNs show strong robustness to bit flip errors on the mantissa field:
model accuracy remains unaffected even at a BER as high as $1 \times 10^{-3}$.
This discrepancy is expected, as the mantissa has only a minor influence on the numerical value, while the exponent, which controls power-of-two scaling in FP numbers, has a substantial effect on weight magnitude.
In addition, sign-bit errors are less severe than exponent errors.
The overall accuracy degradation in full FP weights can be partially mitigated by
the mantissa’s resilience.

\subsubsection{{Insights}}
These findings show that the accuracy of FP DNN models is highly vulnerable to bit errors in the exponent and sign fields while remaining largely unaffected by errors in the mantissa.
Thus, we believe that, \textbf{it is feasible to safeguard only the exponent and sign bits}.
This strategy can eliminate the need for full-bit protection while significantly reducing resource overhead.

One may consider reusing existing ECC schemes developed for INT CIMs~\cite{etcim, mac-ecc, successive}, which reduce overhead by protecting only MAC results via pre-accumulation or pre-encoding, rather than safeguarding individual bit cells. However, as discussed in Section~\ref{fp_operation}, FP MACs require bit-level protection due to operations like exponent max-finding and differencing, making INT-focused ECCs incompatible. While traditional storage ECCs offer bit-wise protection, they introduce substantial overhead. For example, applying a Hamming code to the 6-bit sign and exponent field of each FP16 weight requires 5 redundant bits---an 83.3\% increase in SRAM usage---plus significant logic overhead. These challenges underscore the need for a resource-efficient ECC scheme tailored to FP CIM. The key question remains: \textbf{how can we design an ECC that provides robust protection without excessive cost?}

\subsection{Identifying Optimal Protection Scheme: One4N ECC}
\label{sec:ecc}

\begin{figure}[!t]
\centering
\includegraphics[width=1.0\linewidth]{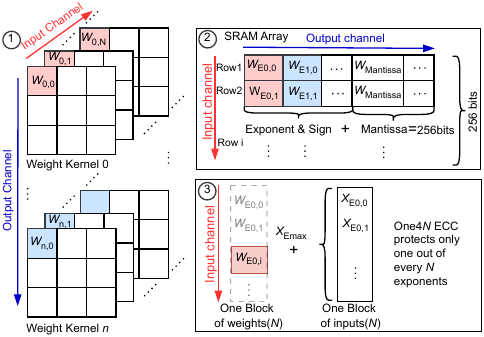}
\caption{Block partitioning of weight kernels \& One4N ECC.}
\vskip -6pt
\label{blocks}
\end{figure}

\begin{figure*}[!t]
\centering
\includegraphics[width=1.0\textwidth]{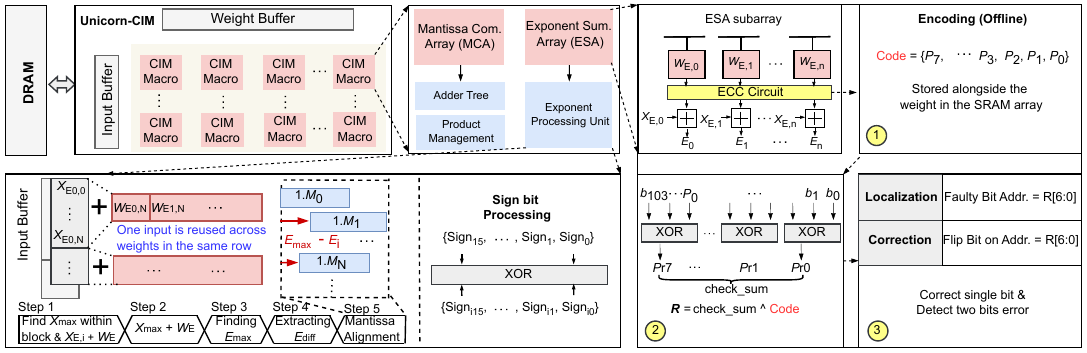 }
\caption{Proposed Unicorn-CIM macro architecture. One4N ECC is embedded into the processing of the exponent and sign parts.}
\vspace{-6pt}
\label{cim_macro}
\end{figure*}

To develop a tailored efficient protection scheme to safeguard the sign bit and exponent of each FP weight (e.g., FP16), we propose a \textbf{row-based one-for-N (One4N) ECC scheme} within the Unicorn-CIM framework.
This innovative scheme incorporates two key features to maximize hardware implementation efficiency.
\textbf{(1) Unified encoding of exponents}: Exponents for all weights within a single row in a memory array are encoded in a unified manner, significantly reducing redundant bits and streamlining data representation.
\textbf{(2) Selective exponent protection:} Instead of safeguarding the exponent for every weight, the scheme protects only one exponent for every group of $N$ weights whose exponents are enforced to be the same by our algorithm optimization (Section~\ref{sec: co-design}), thereby drastically improving resource efficiency without compromising reliability.
These advantages are explained below.

\subsubsection{{Unified encoding of exponents}} 
It is well known that longer protected data segments require fewer redundant bits per unit.
Taking a $256 \times 256$-bit weight array\footnote{Here, we treat the exponent sum array (ESA) and mantissa multiplication array (MSA) as an entity.} as an example, a CIM macro often computes in parallel along the output channel direction for efficient data reuse, where $16$ weights (16-b each) in a row are multiplied by the same input (Fig.~\ref{blocks}  \tikz[baseline=(char.base)]{
    \node[shape=circle, draw, inner sep=0.5pt] (char) {2};
}). 
By applying unified encoding to protect this row -- comprising \(5 \times 16 = 80\) exponent bits and $16$ sign bits -- a total of $96$ bits must be protected.
Using Hamming code, only $8$ redundant bits are needed, dramatically reducing the number of redundant bits from $96$ to $8$, an $12\times$ improvement in efficiency.
% We evaluate this ECC scheme through digital synthesis.
%zing and simulating the RTL (Verilog) implementation. 
% \textcolor{red}{Our results show that the logic overhead remains at $31.57\%$}.

\subsubsection{{Selective exponent protection}}  
Our One4N ECC scheme groups weights along the input channel direction (Fig.~\ref{blocks}  \tikz[baseline=(char.base)]{
    \node[shape=circle, draw, inner sep=0.5pt] (char) {1};
}) and assumes that all \(N\) weights in a group share the same exponent value. 

Thus, only a unique exponent per group needs to be stored and protected (Fig.~\ref{blocks}  \tikz[baseline=(char.base)]{
    \node[shape=circle, draw, inner sep=0.5pt] (char) {3};
}).
% Once errors in this single exponent are detected and corrected, the corrected value is transferred to the exponent computation unit for operations such as addition and determining the maximum value.
This strategy reduces the number of protected exponents by a factor of \(N\), substantially decreasing resource requirements and improving overall efficiency.
Thus, for an $N \times 256$ bit-block in a $256 \times 256$-bit weight array, where each row stores sixteen FP16 weights, the total number of bits (TB) required for protection is expressed as:
\begin{equation}
\text{TB} = 5 \times 16 + N \times 16.
\label{eq:tb}
\end{equation}
Yet, to unleash the full potential of this row-based One4N scheme, the algorithm-hardware co-design is necessary as not all \(N\) weights in a group share the same exponent in a general FP DNN model.

\begin{figure}[!t]
\centering
\includegraphics[width=1.0\linewidth]{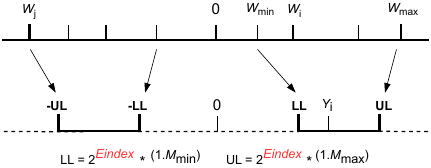}
\caption{Illustration of the weight rescale in the proposed One4N ECC. \( M_{\min} \) and \( M_{\max} \) represent the minimum and maximum values that the mantissa can express.}
\vskip -6 pt
\label{rescale}
\end{figure}

\subsection{Algorithm-Hardware Co-Design}
\label{sec: co-design}

\subsubsection{{Algortihm optimzation}}

Unicorn-CIM leverages an algorithm-level optimization technique to facilitate the exponent alignment as required by the One4N ECC protection scheme: grouping every \(N\) weight along the input channel of the kernels to form a block\footnote{Any remaining weights (i.e., $<N$) are grouped into an additional block.} as shown in Fig.~\ref{blocks} \tikz[baseline=(char.base)]{
    \node[shape=circle, draw, inner sep=0.5pt] (char) {1};
}. 
We then fine-tune a pre-trained FP DNN model by enforcing the same exponent for all weights in a block. 
This strategy is similar to quantization-aware training, but it only applies to the exponent part of FP weights.

Specifically, within each block, we extract all the exponents, sort them in descending order, and select the exponent \(E_{\text{index}}\) at a designated position, indicated by the index.
This selected \(E_{\text{index}}\) is then used to calculate the range of floating-point values it can represent, with the \text{Lower Limit (LL)} and \text{Upper Limit (UL)} defined accordingly (for negative weights, denoted as \((- \text{UL}, - \text{LL})\)). 
Subsequently, the original weights within the block are rescaled to fit within this calculated range, ensuring that all values share the same exponent, as shown in Fig.~\ref{rescale}.
For instance, considering positive weights, the data is rescaled as below
\begin{equation}
Y_i = \left( \frac{W_i - W_{\text{Min}}}{W_{\text{Max}} - W_{\text{Min}}} \right) \times (\text{UL} - \text{LL}) + \text{LL}.
\label{eq:normalization}
\end{equation}
Here, $W_{\text{Max}}$ represents the maximum value of the positive weights, while $W_{\text{Min}}$ denotes their minimum value.
During fine-tuning, our training methodology involves determining the exponent(\(E_{\text{index}}\)) for each block from the pre-trained weights and then keeping this exponent fixed throughout training while only updating mantissa parts.
Consequently, the range \((\text{LL}, \text{UL})\) remains constant.
Our comprehensive evaluations in Section~\ref{performance_N} show that this strategy works well with negligible accuracy loss.

\begin{table*}[!t]
\centering

\caption{
\textbf{} Fine-tuning accuracy ratios of various models with different \(N\) sizes and index values to retrained baseline\textsuperscript{1}.}
\vskip -3pt
\label{sweep}
\scalebox{1}{
    \begin{threeparttable}
    \begin{tabular}{|c|cccc|cccc|cccc|}
    \hline
    Size of N & \multicolumn{4}{c|}{4} & \multicolumn{4}{c|}{8} & \multicolumn{4}{c|}{16} \\ \hline
    Index     & \multicolumn{1}{c|}{1st} & \multicolumn{1}{c|}{2nd} & \multicolumn{1}{c|}{3rd} & 4th & \multicolumn{1}{c|}{1st} & \multicolumn{1}{c|}{2nd} & \multicolumn{1}{c|}{3rd} & 4th & \multicolumn{1}{c|}{1st} & \multicolumn{1}{c|}{2nd} & \multicolumn{1}{c|}{3rd} & 4th \\ \hline
    ResNet18     & \multicolumn{1}{c|}{88.91\%} & \multicolumn{1}{c|}{89.44\%} & \multicolumn{1}{c|}{87.98\%} & 77.59\% & \multicolumn{1}{c|}{95.3\%} & \multicolumn{1}{c|}{\textbf{99.22\%}} & \multicolumn{1}{c|}{98.97\%} & 94.57\% & \multicolumn{1}{c|}{95.16\%} & \multicolumn{1}{c|}{98.65\%} & \multicolumn{1}{c|}{97.21\%} & 97.11\% \\ \hline
    nnUNet   & \multicolumn{1}{c|}{83.75\%} & \multicolumn{1}{c|}{85.94\%} & \multicolumn{1}{c|}{83.65\%} & 76.77\% & \multicolumn{1}{c|}{94.51\%} & \multicolumn{1}{c|}{\textbf{99.14\%}} & \multicolumn{1}{c|}{98.17\%} & 93.68\% & \multicolumn{1}{c|}{94.77\%} & \multicolumn{1}{c|}{97.11\%} & \multicolumn{1}{c|}{97.01\%} & 96.52\% \\ \hline
    Yolov5    & \multicolumn{1}{c|}{84.15\%} & \multicolumn{1}{c|}{86.77\%} & \multicolumn{1}{c|}{83.84\%} & 75.34\% & \multicolumn{1}{c|}{95.63\%} & \multicolumn{1}{c|}{99.10\%} & \multicolumn{1}{c|}{\textbf{99.19\%}} & 96.45\% & \multicolumn{1}{c|}{94.68\%} & \multicolumn{1}{c|}{98.29\%} & \multicolumn{1}{c|}{98.18\%} & 97.35\% \\ \hline
    TinyViT   & \multicolumn{1}{c|}{85.88\%} & \multicolumn{1}{c|}{86.82\%} & \multicolumn{1}{c|}{84.35\%} & 73.27\% & \multicolumn{1}{c|}{95.3\%} & \multicolumn{1}{c|}{\textbf{98.53\%}} & \multicolumn{1}{c|}{98.45\%} & 95.96\% & \multicolumn{1}{c|}{94.22\%} & \multicolumn{1}{c|}{96.78\%} & \multicolumn{1}{c|}{96.88\%} & 95.49\% \\ \hline
    \end{tabular}
    \end{threeparttable}
    }
    \label{tab::compare}
    \vspace{-1mm}
\begin{flushleft}
\parbox{\linewidth}{\hspace{5mm} \small Retrained baseline\textsuperscript{1}: ResNet18 69.7\%, nnUNet 85.2\%, Yolov5 62.4\%, TinyViT 83.1\%.}
\end{flushleft}
\vskip -6pt
\end{table*}

\renewcommand{\arraystretch}{1}

\subsubsection{{Hardware development}}

To accommodate the algorithm optimization, Unicorn-CIM integrates minimal hardware resources (i.e., an ECC circuit), into the general FP CIM architecture, as shown in Fig.~\ref{cim_macro}. 
We take $N=8$ as an example to illustrate the hardware development and processing flow in detail.
For \(N = 8\) (i.e., a $8\times 256$-bit block), we store and protect only a single row of exponents along with all the sign bits, resulting in a total of $208$ bits according to Eq.~\eqref{eq:tb}. 
These bits are divided into two rows for encoding, with each requiring 8 bits. 
The redundant bits generated by the Hamming code are appended to the exponent and stored together within each row of the CIM.
% The corresponding redundant bits (hamming code) are stored alongside the data within a row in the CIM.

We inserted the ECC circuit between the Exponent Summation Array (ESA) and the adder. Before performing exponent summation, the exponent data is first sent to the ECC circuit for bit error detection and correction. The Hamming code is pre-encoded offline and stored in the ESA array along with the exponent data, as shown in the (Fig.~\ref{cim_macro} \tikz[baseline=(char.base)]{
    \node[shape=circle, draw, inner sep=0.5pt] (char) {1};
}).
The stored data and hamming code as is re-encoded in the same manner to obtain a new checksum. 
This checksum is then XORed with the original Hamming code to produce an 8-bit error syndrome \(R\), as illustrated in (Fig.~\ref{cim_macro} \tikz[baseline=(char.base)]{
    \node[shape=circle, draw, inner sep=0.5pt] (char) {2};
}). 
(Fig.~\ref{cim_macro} \tikz[baseline=(char.base)]{
    \node[shape=circle, draw, inner sep=0.5pt] (char) {3};
}) shows the process of bit error detection and correction
(i) If \(R\) is all zeros, no error has occurred.  
(ii) If \(R[7]\) is 1, it indicates a single-bit error, and the position of the error is specified by \(R[6:0]\). 
The erroneous bit can be corrected by flipping the bit at the indicated position.  
(iii) If \(R[7]\) is 0 but \(R[6:0]\) is non-zero, it indicates the occurrence of two or more bit errors, which cannot be corrected by the Hamming code. 
Since the probability of multiple-bit errors occurring within a single row is very low, this ECC scheme provides sufficient protection. 

The corrected data is then sent to the {Exponent Processing Unit}. 
Since only one exponent is stored for a group of $N$ exponents, the required {SRAM cell count} is significantly reduced.
To minimize data movement, weight data is stored along the row direction, aligning with the {output channel direction}. 
Each weight in a row is added to the {same input data}. 
The exponent calculation process consists of five steps:
\textbf{Step 1: Maximum Value Selection}.  
From every $N$ input values, the maximum input exponent (\(X_{\text{max}}\)) is selected. 
At the same time, all $N$ inputs are added to their corresponding shared weight exponents (\(X_{\text{E,i}}\) + \(W_{\text{E}}\)), forming sums that will be used for later exponent difference calculations. 
\textbf{Step 2: Computing the Maximum Exponent Sum}.  
The maximum values from multiple input groups are added to their corresponding weight exponents (\(X_{\text{max}}\) + \(W_{\text{E}}\)).
\textbf{Step 3: Determining the Overall Maximum Value}.  
The overall maximum (\(E_{\text{max}}\)) is determined by finding the maximum exponent sum among all groups obtained in Step 2.
\textbf{Step 4: Difference Calculation}.  
The maximum exponent sum is subtracted from all previously computed sums of inputs and weight exponents, resulting in exponent difference (\(E_{\text{diff}}\)).
\textbf{Step 5: Mantissa Alignment}.  
The mantissa is shifted and aligned based on the \(E_{\text{diff}}\).
The aligned mantissa is then sent to the Mantissa Multiplication Array to complete the subsequent multiplication operations.

During FP multiplication, the sign bit requires only a simple XOR operation. 
As such, the sign bits are routed to the Sign Processing Unit (XOR array), where the processed sign is combined with the product of the mantissas to generate the final output. 
The Unicorn-CIM macro reduces the data storage requirements for exponents and the number of adders needed, enhancing efficiency and resource utilization.

\subsubsection{{Discussions}}
The proposed exponent alignment is compatible with various ECC encoding schemes, offering flexibility in addressing diverse error correction needs. 
For instance, BCH codes \cite{chen2007types} can be used for multi-bit error correction, though they come with higher resource demands. 
Additionally, our exponent alignment training scheme enables on-chip training protection. 
By fixing exponents during training, the system minimizes the impact of bit errors on the weight values, thereby enhancing the robustness of the overall architecture.
Some prior works \cite{fpcim2024_1, fpcim2022} improve the inference efficiency of FP CIM by performing offline weight pre-alignment, which appears to enhance resilience without the need for exponent protection. 
However, sign-bit protection is still necessary, and pre-alignment introduces additional errors, even with a 5-bit extension of each mantissa. 
This significantly increases the energy consumption of multiplication operations. Moreover, this method is not suitable for on-chip training, limiting its applicability in scenarios that require adaptability and fine-tuning.

 \section{Experiment Set-up and Evaluations}

\subsection{Experiment Set-up}

To evaluate Unicorn-CIM, we fine-tune FP DNNs on a workstation equipped with four NVIDIA A6000 GPUs and an AMD EPYC 7313 CPU. 
The selected model benchmarks include ResNet18, YOLOv5, nnUNet, and TinyViT, as presented in Table \ref{models}. 
The benchmark models cover different tasks, such as image
classification, object detection, object segmentation, and different
network architectures such as convolution layers and transformer
architecture.
For hardware evaluation, we implemented and assessed our row-based One4N ECC scheme using Cadence tools for simulation and synthesis. We use the TSMC's N16 ADFP (Academic Design Foster Package) for design. 
The Exponent Processing Unit without ECC protection was chosen as the hardware baseline.

\begin{table}[!tb]
\centering
\caption{Summary of DNN models, datasets, and BER range.}
\vskip -3pt
\label{models} % Place label immediately after caption
\begin{tabular}{|c|c|c|c|}
\hline
DNN Model & Dataset  & BER range                         & Num. per BER/Total      \\ \hline
ResNet18     & ImageNet & \multirow{4}{*}{{[}1E-8, 1E-2{]}} & \multirow{4}{*}{100/24K} \\ \cline{1-2}
Yolov5    & COCO     &                                   &                         \\ \cline{1-2}
nnUNet   & KiTS19   &                                   &                         \\ \cline{1-2}
TinyViT   & ImageNet &                                   &                         \\ \hline
\end{tabular}
% \vskip -6pt
\end{table}

\subsection{Evaluations}
\label{performance_N}

\subsubsection{{Algorithm optimization}}

We conducted a comprehensive exploration of various combinations of $N$ and index values to evaluate the effectiveness of the proposed fine-tuning algorithm in Section~\ref{sec: co-design}. 
The results for all benchmark models are summarized in Table~\ref{tab::compare}.
We observe that setting $N = 8$ offers the best trade-off, preserving over $99\%$ of baseline accuracy (i.e., without exponent alignment). While $N = 16$ incurs moderate degradation, $N = 4$ significantly impacts performance due to fewer data points, making the distribution more sensitive to outliers during exponent alignment. Conversely, $N = 16$ suffers from excessive grouping, constraining weight flexibility and reducing accuracy.
This highlights a trade-off: different $N$ values can be selected based on application needs to balance efficiency and accuracy. Regarding exponent index selection, using the 2nd or 3rd largest exponent yields optimal results, whereas the 1st and 4th degrade performance—index 1 often inflates weights, while index 4 underrepresents them. Thus, $N = 8$ with index 2 or 3 ensures optimal quantization performance.

\subsubsection{FP DNN resilience with tailored protection}

\begin{figure}[!]
\centering
\begin{subfigure}{0.24\textwidth}
    \centering
    \includegraphics[width=\linewidth]{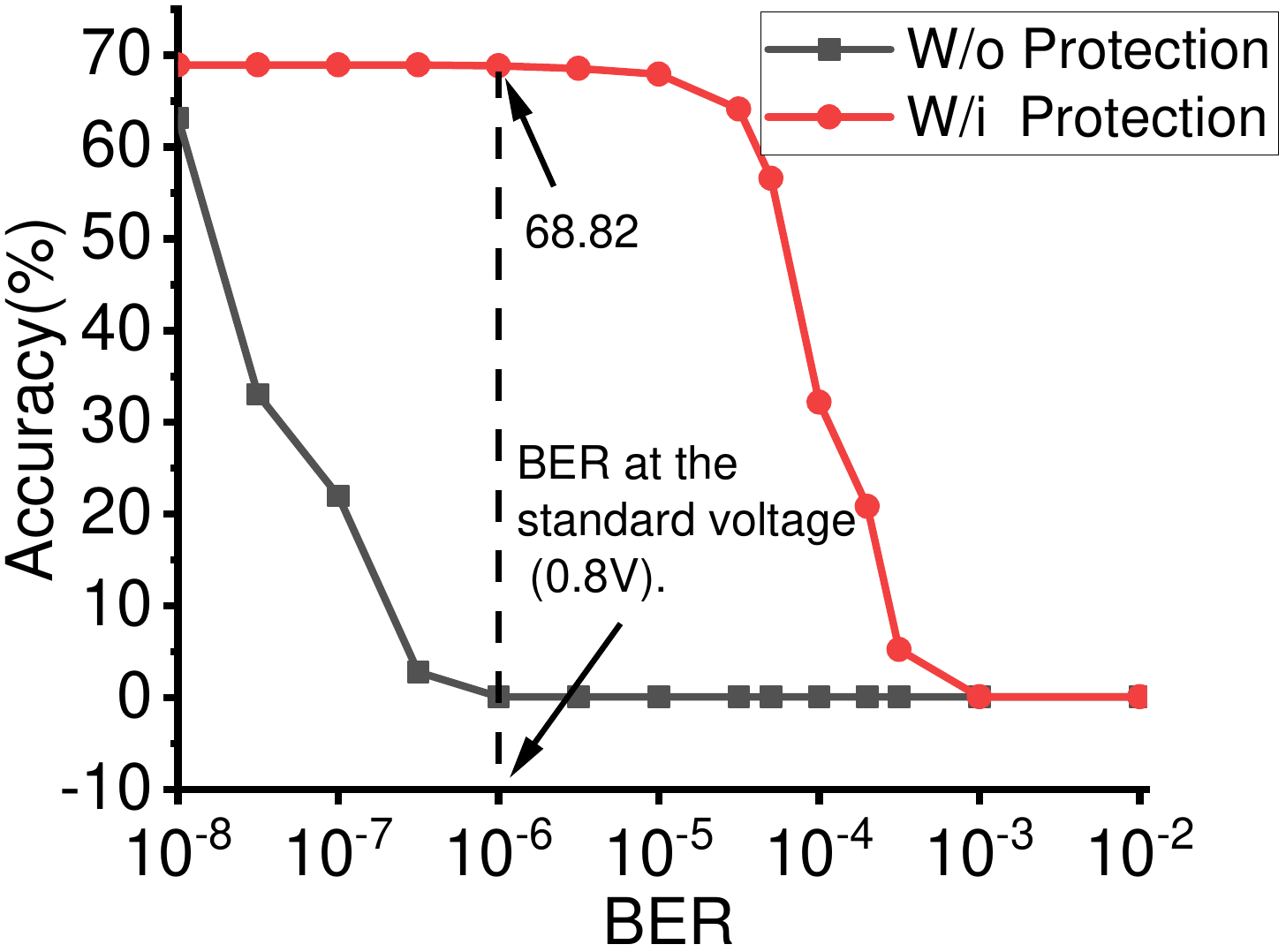}
    \caption{ResNet18}
    \label{fig:fp16_1}
\end{subfigure}
\hfill
\begin{subfigure}{0.24\textwidth}
    \centering
    \includegraphics[width=\linewidth]{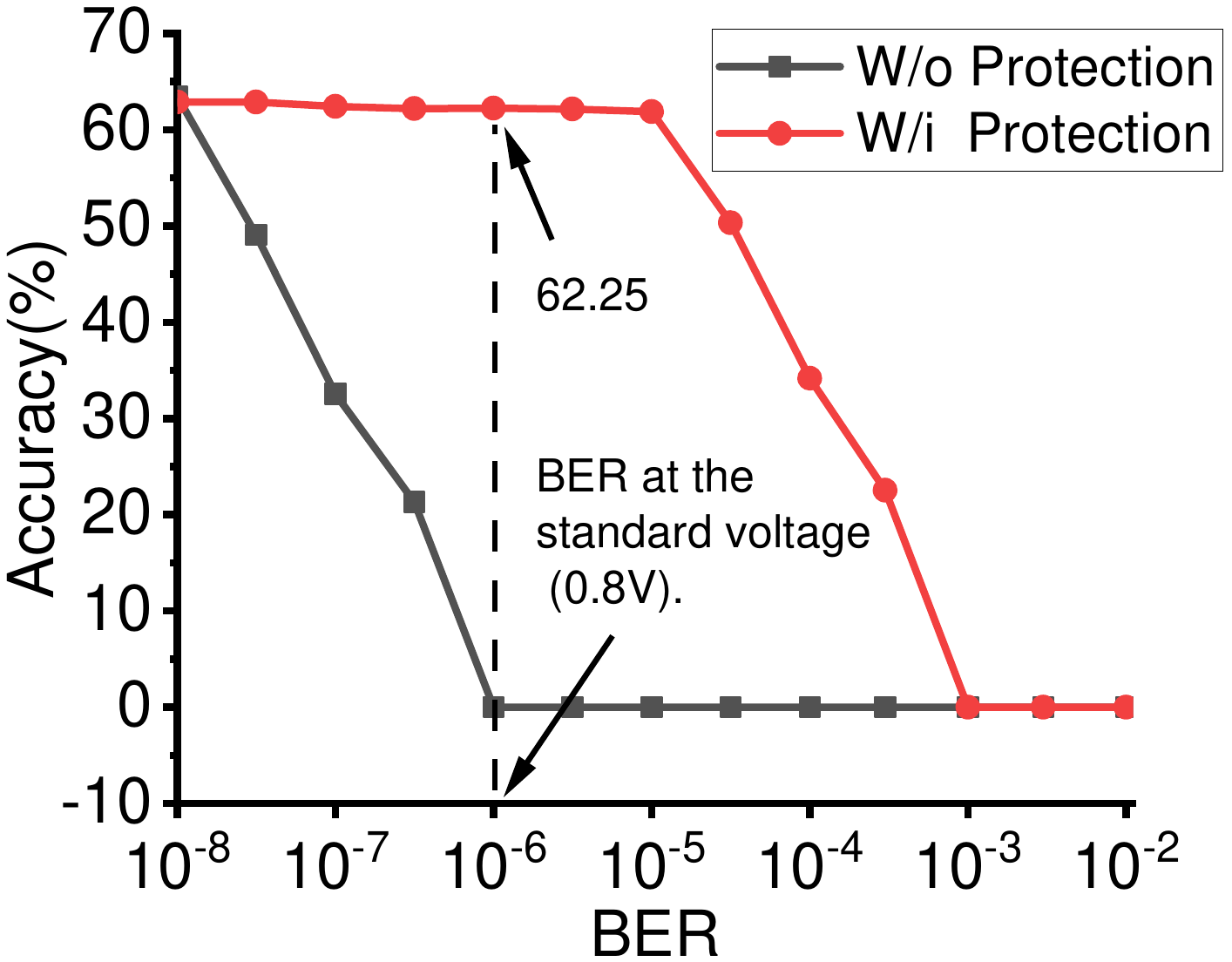}
    \caption{Yolov5}
    \label{fig:fp16_2}
\end{subfigure}
\vskip\baselineskip
\begin{subfigure}{0.24\textwidth}
    \centering
    \includegraphics[width=\linewidth]{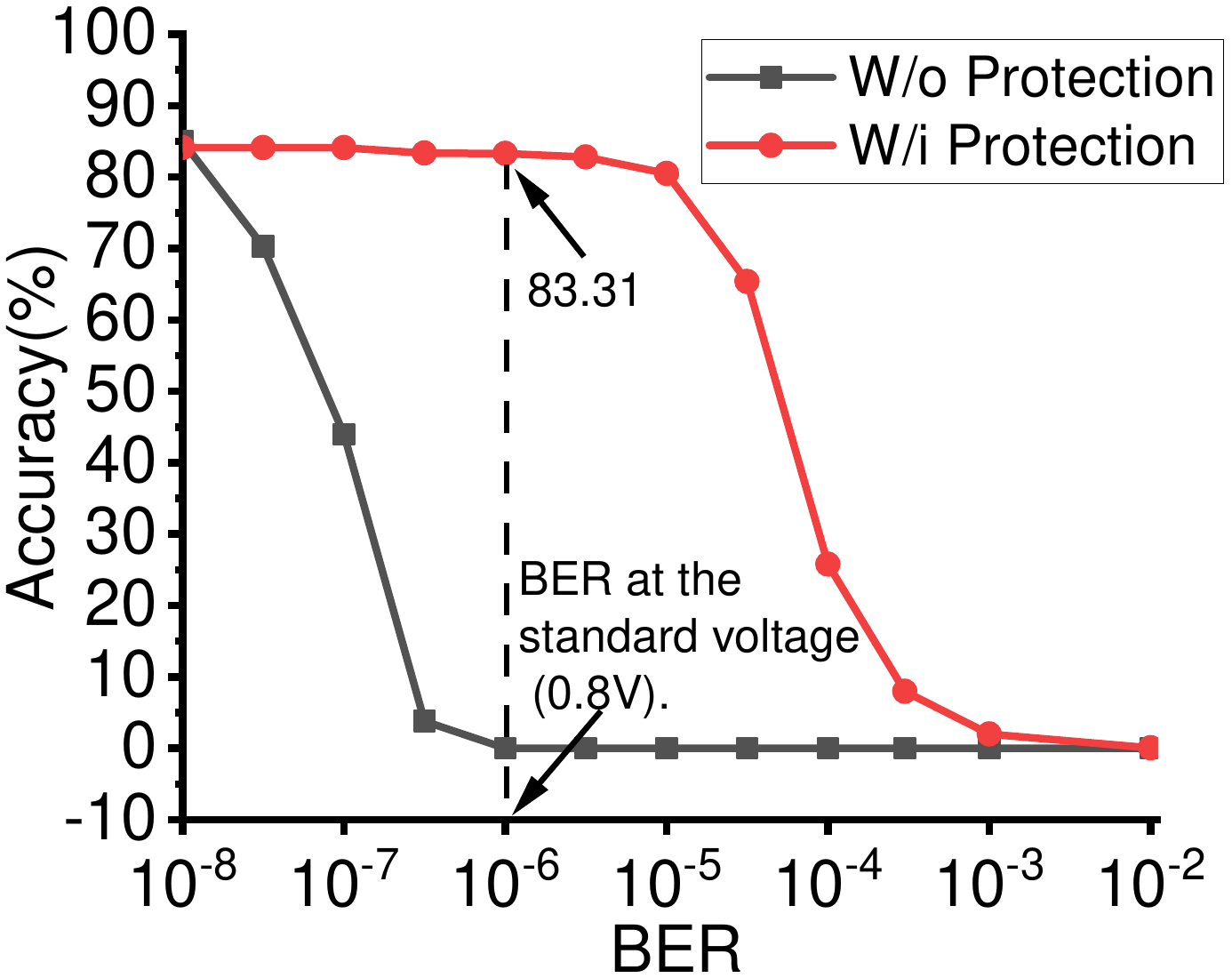}
    \caption{nnUNet}
    \label{fig:fp16_3}
\end{subfigure}
\hfill
\begin{subfigure}{0.24\textwidth}
    \centering
    \includegraphics[width=\linewidth]{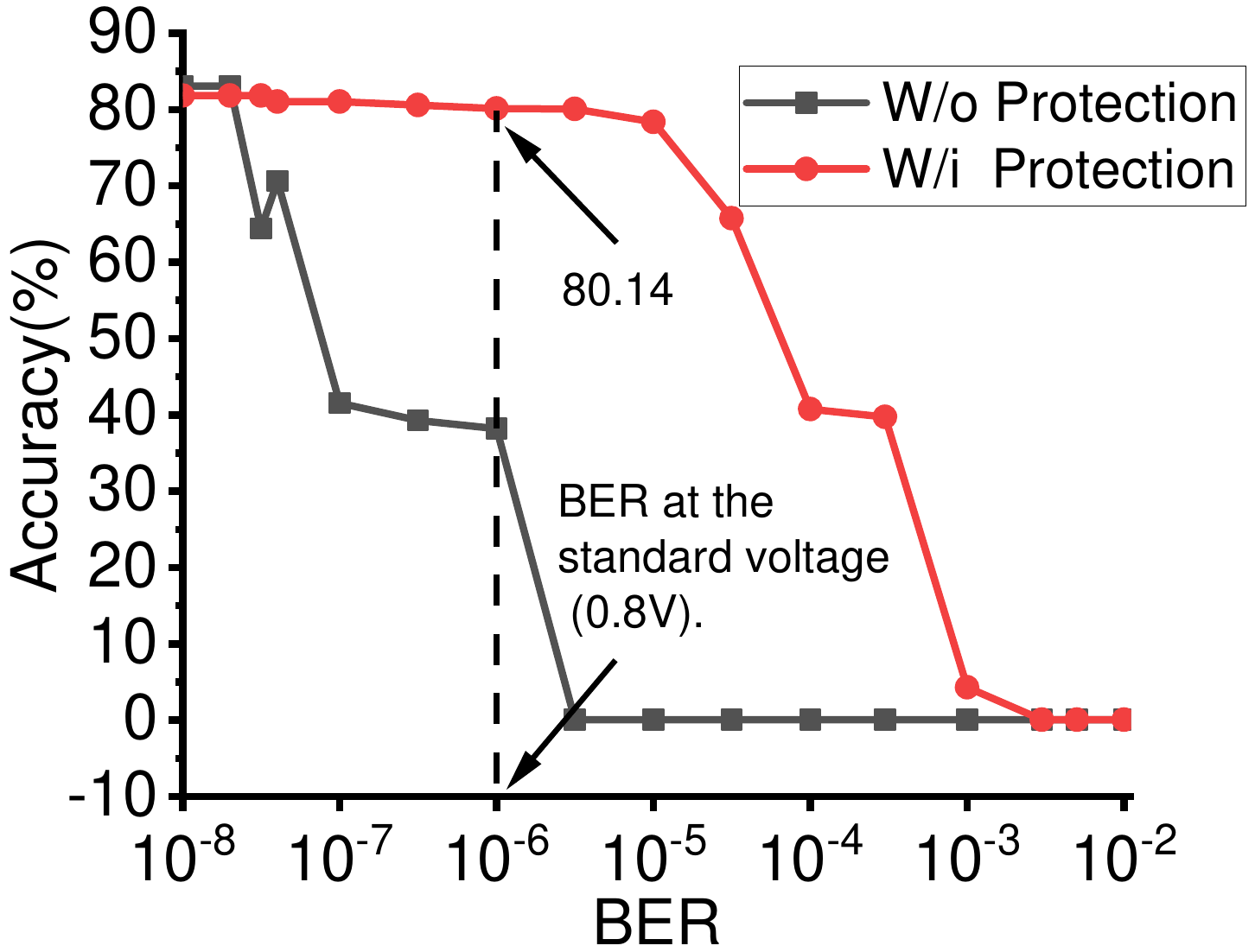}
    \caption{TinyViT}
    \label{fig:fp16_4}
\end{subfigure}

\caption{Accuracy vs. BER Wi/o ECC protection on benchmarking DNN models.}
\label{fig:ecc_after}
\end{figure}

We conducted experiments to evaluate the performance of the row-based One4N ECC. Accuracy was measured under various BER conditions, ranging from \(1 \times 10^{-8}\) to \(1 \times 10^{-2}\). 
For each BER, $1000$ experiments were performed using exponent-aligned model weights trained with our algorithm, under both protected and unprotected conditions, and the average accuracy was recorded.
The results, presented in Fig.~\ref{fig:ecc_after}, show that even at a BER of \(10^{-6}\), which corresponds to the standard operating voltage (0.8$V$), the model accuracy without ECC protection degrades drastically, approaching zero.
However, with the protection of our One4N ECC, the models maintain a high level of accuracy.

\begin{figure}[!t]
    \centering
\begin{subfigure}{0.24\textwidth}
    \centering
    \includegraphics[width=\linewidth]{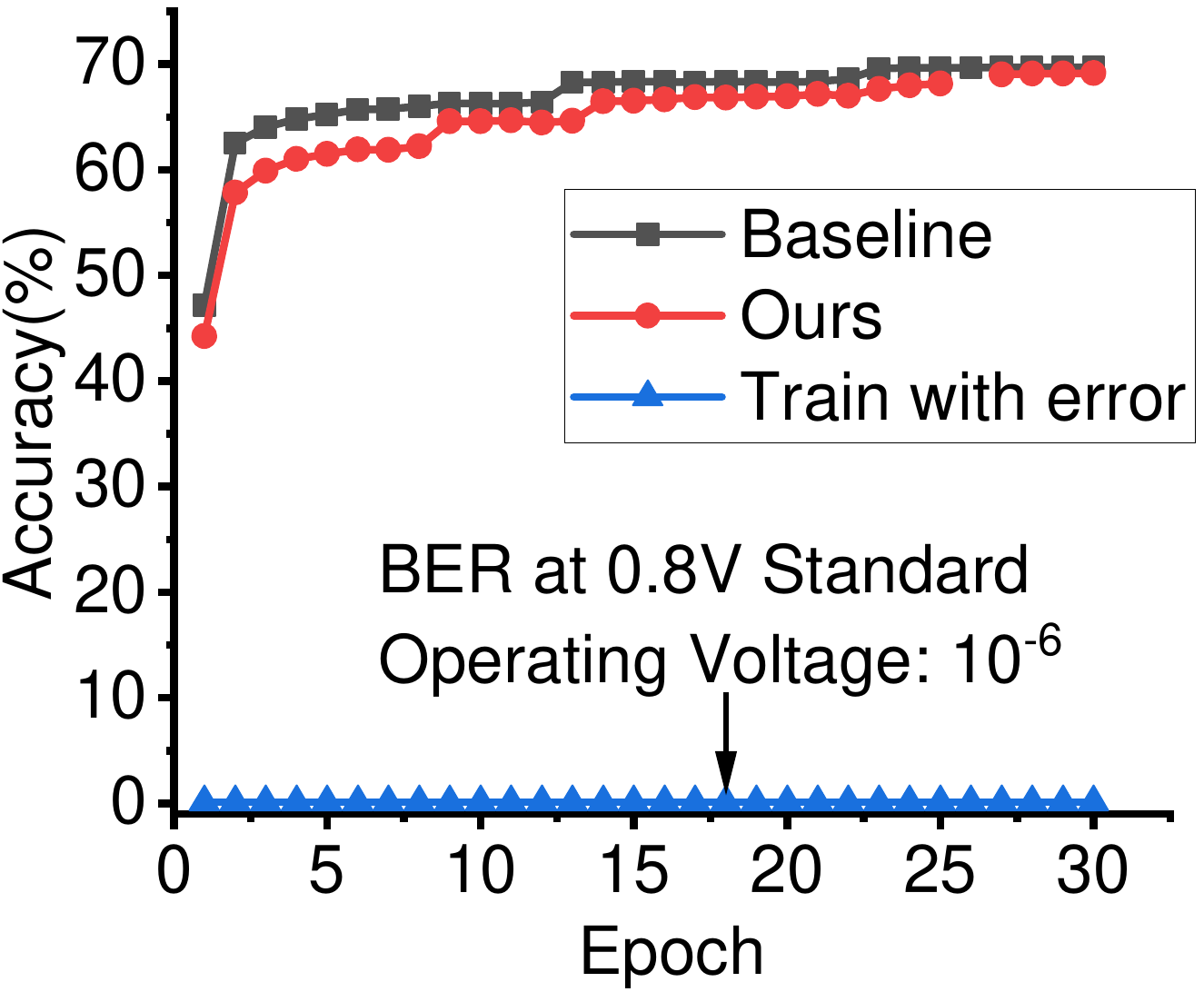}
    \caption{ResNet18}
    \label{fig:fp17}
\end{subfigure}
\hfill
\begin{subfigure}{0.24\textwidth}
    \centering
    \includegraphics[width=\linewidth]{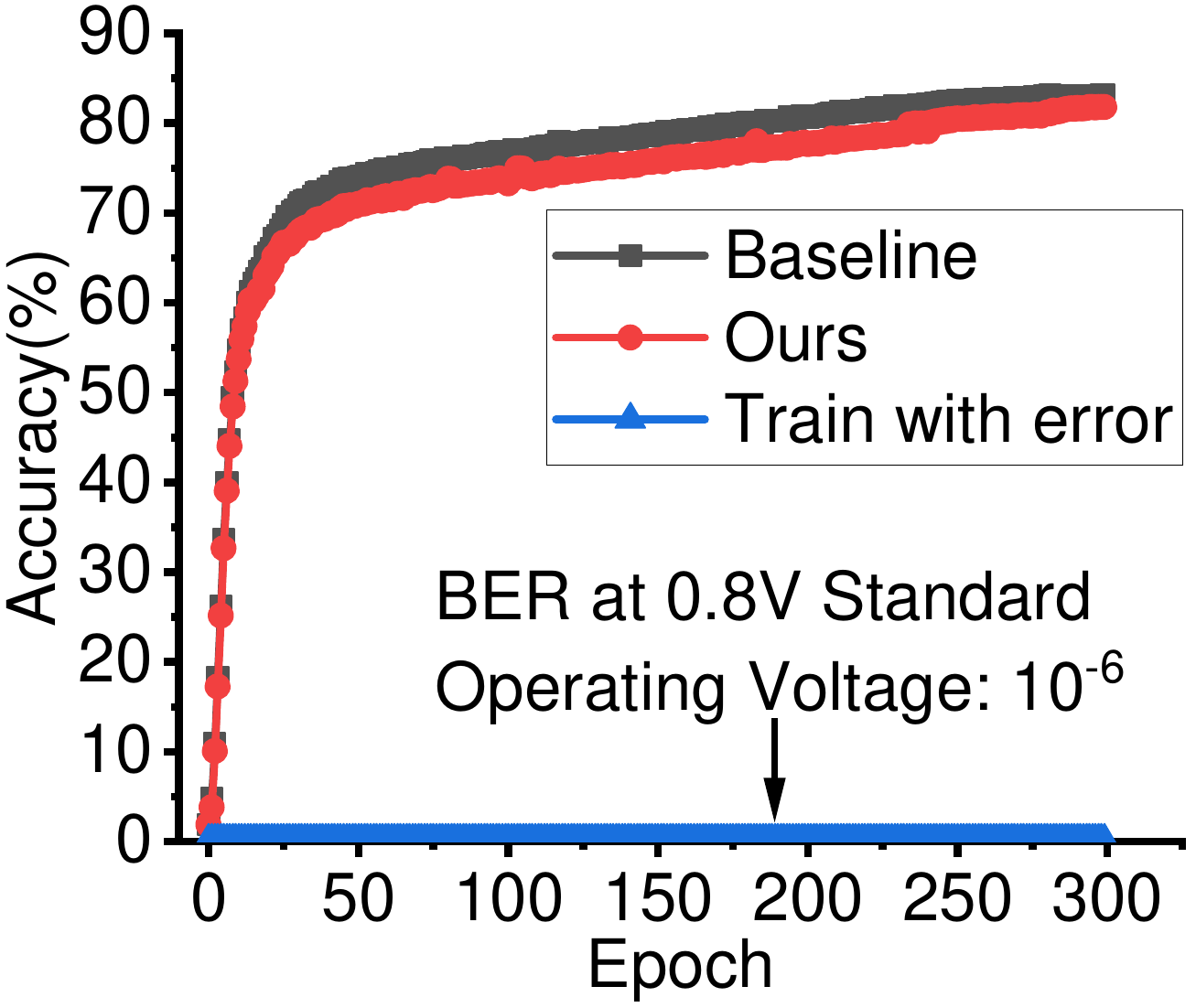}
    \caption{TinyViT}
    \label{fig:fp18}
\end{subfigure}
    \caption{Illustration of the training process Wi/o ECC protection.}
    \label{train}
\end{figure}

We also recorded the training process under different error injection conditions, dividing the experiments into three groups: (1) training without error injection, (2) training with error injection under a \(10^{-6}\) BER, and (3) training with error injection while applying our exponential alignment fine-tuning method. 
We use the ResNet18 and TinyVit models as examples for our study, and the results are illustrated in Fig.~\ref{train}.  
The comparisons show that our fine-tuning method does not increase the number of epochs required for convergence. 
When error injection is introduced, the baseline model frequently encounters NaN issues, leading to training failures—most likely due to bit flips in the exponent field causing numerical overflows or invalid operations during computation.
In contrast, our method effectively mitigates these issues and maintains high accuracy.  
\textbf{Since FP CIM is primarily designed for high-precision, on-device training, our findings demonstrate that without proper protection mechanisms, achieving the desired performance is infeasible}.

\subsubsection{Hardware efficiency}

To evaluate the efficiency of our Unicorn-CIM macro, we conduct comparisons using a \(256 \times 256\) SRAM array and various ECC protection schemes. 
The data, summarized in Table~\ref{tab::overhead}, highlights the improvements of our method. 
For full-bit protection, where both exponent and mantissa are safeguarded, separate encoding is required as they belong to different modules. 
This results in the mantissa requiring an additional five redundant bits, leading to a total of $40,960$ redundant bits. 
In contrast, protecting only the exponent and sign reduces this to $20,480$ bits. 
Our Unicorn-CIM macro requires just $512$ redundant bits, achieving improvements of $80\times$ and $40\times$, respectively. 
For logic overhead,  we use the Exponent Processing Unit as the baseline. 
Unicorn-CIM macro demonstrates $8.3\times$ and $3.51\times$ improvements, respectively. 
Additionally, since the Unicorn-CIM macro stores and protects only one exponent for every group of \(N\) exponents, the number of SRAM bit cells required for exponents is just $2560$, achieving an \(8\times\) reduction (for \(N=8\)).

Taking the power consumption of the Exponent Processing Unit without ECC as the baseline, it accounts for approximately $40\%$ of the total power consumption of a macro\cite{FP_energy}.
According to simulation results, our proposed method reduces the power consumption of the traditional ECC scheme from \(12.55\%\) of the baseline to \(3.69\%\). 
As a result, the overall power overhead for the full operation of the FP CIM accelerator is only 
1.48\%. 

\subsubsection{Summarization}

Our Unicorn-CIM introduces only a small amount of additional logic resources while providing effective protection with negligible impact on accuracy, making it suitable for high-precision inference and training scenarios. 
This work paves the way for developing more reliable and efficient CIM
hardware.
For future work, we will extend our research to DNN models with FP8 precision.
We will also explore new fine-tuning schemes (e.g., ) that enable more efficient ECC.

\begin{table}[!t]
\centering
\caption{\textbf{} Hardware efficiency comparisons.}
\begin{threeparttable}
\resizebox{\linewidth}{!}{%
\begin{tabular}{|l|c|c|c|}
\hline
\multicolumn{1}{|c|}{Array of 256×256}                                         & \begin{tabular}[c]{@{}c@{}}ECC\\ Redundant bits\end{tabular} & \begin{tabular}[c]{@{}c@{}}Logic\\ Overhead$^{1}$\end{tabular} & \begin{tabular}[c]{@{}c@{}}SRAM bit cell\\ for exponent\end{tabular} \\ \hline
\begin{tabular}[c]{@{}l@{}}Traditional ECC\\ for Entire FP Num.\end{tabular}   & 40960(80×)                                                   & 74.44\% (8.3×)                                           & \multirow{3}{*}{20480(8×)}                                           \\ \cline{1-3}
\begin{tabular}[c]{@{}l@{}}Traditional ECC\\ for Exponent \& Sign\end{tabular} & 20480(40×)                                                   & 31.55\%(3.51×)                                           &                                                                      \\ \cline{1-3}
\begin{tabular}[c]{@{}l@{}}Row-based ECC\\ for Entire FP Num.\end{tabular}     & 4352(8.5×)                                                   & 73.64\%(8.2×)                                            &                                                                      \\ \hline
Ours                                                                           & \textbf{512}                                                 & \textbf{8.98\%}                                          & \textbf{2560}                                                        \\ \hline
\end{tabular}
}% end resizebox
\begin{tablenotes}
    \footnotesize
    \item[1] Baseline is the area of Exponent Processing Unit.
\end{tablenotes}
\end{threeparttable}
\label{tab::overhead}
\end{table}

\section{Conclusion}

This paper has studied the vulnerability of FP CIM, which is essential to accelerate high-precision DNNs in both inference and on-device training scenarios.
Our tailored fault injection experiments identify that the exponent and sign bits are highly sensitive to bit errors, while the mantissa is largely error-tolerant.
With this key insight, we propose One4N ECC, a lightweight algorithm-hardware co-design framework to protect FP CIM.  
Extensive evaluations demonstrate that when applied to a $256 \times 256$-bit weight array under the BER corresponding to the standard operating voltage, our ECC scheme maintains model accuracy with minimal impact while incurring only $3.69\%$ power overhead and $8.98\%$ logic overhead--using the Exponent Processing Array as the baseline--and reducing the SRAM bit cell for exponents by $8\times$.
Our work opens the door to the development of trustworthy and efficient CIM hardware.

% \begin{figure}[htbp]
% \centerline{\includegraphics{fig1.png}}
% \caption{Example of a figure caption.}
% \label{fig}
% \end{figure}

% \section*{Acknowledgment}

% The preferred spelling of the word ``acknowledgment'' in America is without 
% an ``e'' after the ``g''. Avoid the stilted expression ``one of us (R. B. 
% G.) thanks $\ldots$''. Instead, try ``R. B. G. thanks$\ldots$''. Put sponsor 
% acknowledgments in the unnumbered footnote on the first page.

% \section*{References}

\vspace{12pt}


\begin{thebibliography}{00}

\bibitem{survey}
W.~Zhang, B.~Gao, J.~Tang, P.~Yao, S.~Yu, M.-F. Chang, H.-J. Yoo, H.~Qian, and H.~Wu, ``Neuro-inspired computing chips,'' \emph{Nature Electronics}, vol.~3, pp. 371--382, 07 2020.

\bibitem{intcimv1}
J.-H. Yoon, M.~Chang, W.-S. Khwa, Y.-D. Chih, M.-F. Chang, and A.~Raychowdhury, ``29.1 a 40nm 64kb 56.67tops/w read-disturb-tolerant compute-in-memory/digital rram macro with active-feedback-based read and in-situ write verification,'' in \emph{2021 IEEE International Solid-State Circuits Conference (ISSCC)}, vol.~64, 2021, pp. 404--406.

\bibitem{intcimv2}
S.~D. Spetalnick, M.~Chang, B.~Crafton, W.-S. Khwa, Y.-D. Chih, M.-F. Chang, and A.~Raychowdhury, ``A 40nm 64kb 26.56tops/w 2.37mb/mm2rram binary/compute-in-memory macro with 4.23x improvement in density and \>75\% use of sensing dynamic range,'' in \emph{2022 IEEE International Solid-State Circuits Conference (ISSCC)}, vol.~65, 2022, pp. 1--3.

\bibitem{intcimv3}
H.~Fujiwara, H.~Mori, W.-C. Zhao, K.~Khare, C.-E. Lee, X.~Peng, V.~Joshi, C.-K. Chuang, S.-H. Hsu, T.~Hashizume, T.~Naganuma, C.-H. Tien, Y.-Y. Liu, Y.-C. Lai, C.-F. Lee, T.-L. Chou, K.~Akarvardar, S.~Adham, Y.~Wang, Y.-D. Chih, Y.-H. Chen, H.-J. Liao, and T.-Y.~J. Chang, ``34.4 a 3nm, 32.5tops/w, 55.0tops/mm2 and 3.78mb/mm2 fully-digital compute-in-memory macro supporting int12 × int12 with a parallel-mac architecture and foundry 6t-sram bit cell,'' in \emph{2024 IEEE International Solid-State Circuits Conference (ISSCC)}, vol.~67, 2024, pp. 572--574.

\bibitem{fpcim2023}
P.-C. Wu, J.-W. Su, L.-Y. Hong, J.-S. Ren, C.-H. Chien, H.-Y. Chen, C.-E. Ke, H.-M. Hsiao, S.-H. Li, S.-S. Sheu, W.-C. Lo, S.-C. Chang, C.-C. Lo, R.-S. Liu, C.-C. Hsieh, K.-T. Tang, and M.-F. Chang, ``A 22nm 832kb hybrid-domain floating-point sram in-memory-compute macro with 16.2-70.2tflops/w for high-accuracy ai-edge devices,'' in \emph{2023 IEEE International Solid-State Circuits Conference (ISSCC)}, 2023, pp. 126--128.

\bibitem{fpcim2024_2}
W.-S. Khwa, P.-C. Wu, J.-J. Wu, J.-W. Su, H.-Y. Chen, Z.-E. Ke, T.-C. Chiu, J.-M. Hsu, C.-Y. Cheng, Y.-C. Chen, C.-C. Lo, R.-S. Liu, C.-C. Hsieh, K.-T. Tang, and M.-F. Chang, ``34.2 a 16nm 96kb integer/floating-point dual-mode-gain-cell-computing-in-memory macro achieving 73.3-163.3tops/w and 33.2-91.2tflops/w for ai-edge devices,'' in \emph{2024 IEEE International Solid-State Circuits Conference (ISSCC)}, vol.~67, 2024, pp. 568--570.

\bibitem{fpcim2024_1}
T.-H. Wen, H.-H. Hsu, W.-S. Khwa, W.-H. Huang, Z.-E. Ke, Y.-H. Chin, H.-J. Wen, Y.-C. Chang, W.-T. Hsu, C.-C. Lo, R.-S. Liu, C.-C. Hsieh, K.-T. Tang, S.-H. Teng, C.-C. Chou, Y.-D. Chih, T.-Y.~J. Chang, and M.-F. Chang, ``34.8 a 22nm 16mb floating-point reram compute-in-memory macro with 31.2tflops/w for ai edge devices,'' in \emph{2024 IEEE International Solid-State Circuits Conference (ISSCC)}, vol.~67, 2024, pp. 580--582.

\bibitem{fpcim2022}
F.~Tu, Y.~Wang, Z.~Wu, L.~Liang, Y.~Ding, B.~Kim, L.~Liu, S.~Wei, Y.~Xie, and S.~Yin, ``A 28nm 29.2tflops/w bf16 and 36.5tops/w int8 reconfigurable digital cim processor with unified fp/int pipeline and bitwise in-memory booth multiplication for cloud deep learning acceleration,'' in \emph{2022 IEEE International Solid-State Circuits Conference (ISSCC)}, vol.~65, 2022, pp. 1--3.

\bibitem{sun}
X.~Sun, W.~Cao, B.~Crafton, K.~Akarvardar, H.~Mori, H.~Fujiwara, H.~Noguchi, Y.-D. Chih, M.-F. Chang, Y.~Wang, and T.-Y.~J. Chang, ``Efficient processing of mlperf mobile workloads using digital compute-in-memory macros,'' \emph{IEEE Transactions on Computer-Aided Design of Integrated Circuits and Systems}, vol.~43, no.~4, pp. 1191--1205, 2024.


\bibitem{cim2025_2}
W.-S. Khwa, P.-C. Wu, J.-W. Su, C.-Y. Cheng, J.-M. Hsu, Y.-C. Chen, L.-J. Hsieh, J.-C. Bai, Y.-S. Kao, T.-H. Lou, A.~S. Lele, J.-J. Wu, J.-C. Tien, C.-C. Lo, R.-S. Liu, C.-C. Hsieh, K.-T. Tang, and M.-F. Chang, ``14.2 a 16nm 216kb, 188.4tops/w and 133.5tflops/w microscaling multi-mode gain-cell cim macro edge-ai devices,'' in \emph{2025 IEEE International Solid-State Circuits Conference (ISSCC)}, vol.~68, 2025, pp. 1--3.

\bibitem{cao2024addition}
W.~Cao, J.~Gao, X.~Xin, and X.~Zhang, ``Addition is most you need: Efficient floating-point sram compute-in-memory by harnessing mantissa addition,'' in \emph{Proceedings of the 61st ACM/IEEE Design Automation Conference}, 2024, pp. 1--6.

\bibitem{cao_pim}
W.~Cao, Y.~Zhao, A.~Boloor, Y.~Han, X.~Zhang, and L.~Jiang, ``Neural-pim: Efficient processing-in-memory with neural approximation of peripherals,'' \emph{IEEE Transactions on Computers}, vol.~71, no.~9, pp. 2142--2155, 2022.

\bibitem{yi}
Z.~Yi, Y.~Liang, and W.~Cao, ``A hybrid-domain floating-point compute-in-memory architecture for efficient acceleration of high-precision deep neural networks,'' 2025. [Online]. Available: \url{https://arxiv.org/abs/2502.07212}

\bibitem{voltage_scale}
H.~Fujiwara, H.~Mori, W.-C. Zhao, M.-C. Chuang, R.~Naous, C.-K. Chuang, T.~Hashizume, D.~Sun, C.-F. Lee, K.~Akarvardar \emph{et~al.}, ``A 5-nm 254-tops/w 221-tops/mm 2 fully-digital computing-in-memory macro supporting wide-range dynamic-voltage-frequency scaling and simultaneous mac and write operations,'' in \emph{2022 IEEE International Solid-State Circuits Conference (ISSCC)}, vol.~65.\hskip 1em plus 0.5em minus 0.4em\relax IEEE, 2022, pp. 1--3.

\bibitem{dnn_appl}
H.~Hussain, P.~Tamizharasan, and C.~Rahul, ``Design possibilities and challenges of dnn models: a review on the perspective of end devices,'' \emph{Artificial Intelligence Review}, pp. 1--59, 2022.



\bibitem{ber_sram}
F.~Frustaci, M.~Khayatzadeh, D.~Blaauw, D.~Sylvester, and M.~Alioto, ``13.8 a 32kb sram for error-free and error-tolerant applications with dynamic energy-quality management in 28nm cmos,'' in \emph{2014 IEEE International Solid-State Circuits Conference Digest of Technical Papers (ISSCC)}, 2014, pp. 244--245.

\bibitem{etcim}
Y.~Wang, Z.~He, C.~Zhao, Z.~Wu, M.~Gao, H.~Han, S.~Wei, Y.~Hu, F.~Tu, and S.~Yin, ``Etcim: An error-tolerant digital-cim processor with redundancy-free repair and run-time mac and cell error correction,'' in \emph{2024 IEEE Symposium on VLSI Technology and Circuits (VLSI Technology and Circuits)}, 2024, pp. 1--2.

\bibitem{secded}
B.~Crafton, S.~Spetalnick, J.-H. Yoon, W.~Wu, C.~Tokunaga, V.~De, and A.~Raychowdhury, ``Cim-secded: A 40nm 64kb compute in-memory rram macro with ecc enabling reliable operation,'' in \emph{2021 IEEE Asian Solid-State Circuits Conference (A-SSCC)}, 2021, pp. 1--3.

\bibitem{successive}
% \BIBentryALTinterwordspacing
B.~Crafton, Z.~Wan, S.~Spetalnick, J.-H. Yoon, W.~Wu, C.~Tokunaga, V.~De, and A.~Raychowdhury, ``Improving compute in-memory ecc reliability with successive correction,'' in \emph{Proceedings of the 59th ACM/IEEE Design Automation Conference}, ser. DAC '22.\hskip 1em plus 0.5em minus 0.4em\relax New York, NY, USA: Association for Computing Machinery, 2022, p. 745–750. [Online]. Available: \url{https://doi.org/10.1145/3489517.3530526}
% \BIBentrySTDinterwordspacing

\bibitem{pytprchFI}
A.~Mahmoud, N.~Aggarwal, A.~Nobbe, J.~R.~S. Vicarte, S.~V. Adve, C.~W. Fletcher, I.~Frosio, and S.~K.~S. Hari, ``Pytorchfi: A runtime perturbation tool for dnns,'' in \emph{2020 50th Annual IEEE/IFIP International Conference on Dependable Systems and Networks Workshops (DSN-W)}, 2020, pp. 25--31.

\bibitem{tensorFI}
% \BIBentryALTinterwordspacing
Z.~Chen, N.~Narayanan, B.~Fang, G.~Li, K.~Pattabiraman, and N.~DeBardeleben, ``Tensorfi: A flexible fault injection framework for tensorflow applications,'' 2020. [Online]. Available: \url{https://arxiv.org/abs/2004.01743}
% \BIBentrySTDinterwordspacing

\bibitem{ares}
B.~Reagen, U.~Gupta, L.~Pentecost, P.~Whatmough, S.~K. Lee, N.~Mulholland, D.~Brooks, and G.-Y. Wei, ``Ares: A framework for quantifying the resilience of deep neural networks,'' in \emph{2018 55th ACM/ESDA/IEEE Design Automation Conference (DAC)}, 2018, pp. 1--6.

\bibitem{soc_soft_test}
% \BIBentryALTinterwordspacing
Q.~Cheng, Q.~Li, L.~Lin, W.~Liao, L.~Dai, H.~Yu, and M.~Hashimoto, ``How accurately can soft error impact be estimated in black-box/white-box cases? -- a case study with an edge ai soc --,'' in \emph{Proceedings of the 61st ACM/IEEE Design Automation Conference}, ser. DAC '24.\hskip 1em plus 0.5em minus 0.4em\relax New York, NY, USA: Association for Computing Machinery, 2024. [Online]. Available: \url{https://doi.org/10.1145/3649329.3656537}
% \BIBentrySTDinterwordspacing


\bibitem{fpu_2012}
M.~Maniatakos, P.~Kudva, B.~M. Fleischer, and Y.~Makris, ``Low-cost concurrent error detection for floating-point unit (fpu) controllers,'' \emph{IEEE Transactions on Computers}, vol.~62, no.~7, pp. 1376--1388, 2012.

\bibitem{SEU}
Z.~Yan, Y.~Shi, W.~Liao, M.~Hashimoto, X.~Zhou, and C.~Zhuo, ``When single event upset meets deep neural networks: Observations, explorations, and remedies,'' in \emph{2020 25th Asia and South Pacific Design Automation Conference (ASP-DAC)}.\hskip 1em plus 0.5em minus 0.4em\relax IEEE, 2020, pp. 163--168.


\bibitem{mac-ecc}
W.~Li, J.~Read, H.~Jiang, and S.~Yu, ``Mac-ecc: In-situ error correction and its design methodology for reliable nvm-based compute-in-memory inference engine,'' \emph{IEEE Journal on Emerging and Selected Topics in Circuits and Systems}, vol.~12, no.~4, pp. 835--845, 2022.

\bibitem{chen2007types}
S.~Chen, ``What types of ecc should be used on flash memory,'' \emph{Application Note for SPANSION}, 2007.

\bibitem{FP_energy}
F.~Tu, Y.~Wang, Z.~Wu, L.~Liang, Y.~Ding, B.~Kim, L.~Liu, S.~Wei, Y.~Xie, and S.~Yin, ``A 28nm 29.2tflops/w bf16 and 36.5tops/w int8 reconfigurable digital cim processor with unified fp/int pipeline and bitwise in-memory booth multiplication for cloud deep learning acceleration,'' in \emph{2022 IEEE International Solid-State Circuits Conference (ISSCC)}, vol.~65, 2022, pp. 1--3.


\end{thebibliography}
\end{document}